\documentclass[twocolumn]{aa}
\usepackage{graphicx}
\usepackage{txfonts}
\def\mrk{{Mrk~705}}

\def\chandra{{\it Chandra}}

\def\first{{\it FIRST}}
\def\rosat{{\it ROSAT}}

\def\asca{{\it ASCA}}

\def\xte{{\it RXTE}}

\def\et{{et al.\ }}

\def\3c{{3C~273}}
\def\mrk{{Mrk~705}}

\def\ka{{K$\alpha$}}
\def\oiii{{[O~\textsc{iii}]}}

\def\feii{{Fe~\textsc{ii}}}
\def\nh{{N_{\rm H}}}

\def\arcs{{\hbox{$^{\prime\prime}$}}}
\def\deg{^{\circ}}

\def\cm{{\rm\thinspace cm}}
\def\erg{{\rm\thinspace erg}}
\def\eV{{\rm\thinspace eV}}

\def\keV{{\rm\thinspace keV}}
\def\km{{\rm\thinspace km}}
\def\kpc{{\rm\thinspace kpc}}

\def\mJy{{\rm\thinspace mJy}}

\def\pc{{\rm\thinspace pc}}
\def\s{{\rm\thinspace s}}
\def\ks{{\rm\thinspace ks}}
\def\ps{{\rm\thinspace s^{-1}}}

\def\cts{{\rm\thinspace count}}

\def\cps{\hbox{$\cts\s^{-1}\,$}}

\def\ergpscmps{\hbox{$\erg\cm^{-2}\s^{-1}\,$}}

\def\ergps{\hbox{$\erg\s^{-1}\,$}}

\def\kmps{\hbox{$\km\ps\,$}}

\def\pscm{\hbox{$\cm^{-2}\,$}}

\begin{document}
   \titlerunning{An X-ray view of \mrk}
   \authorrunning{L. C. Gallo \et}
   \title{An X-ray view of \mrk}

   \subtitle{A borderline narrow-line Seyfert 1 galaxy}

   \author{L. C. Gallo,
          \inst{1}
          I. Balestra,
          \inst{1}
          E. Costantini,
          \inst{2,3}
          Th. Boller,
          \inst{1}
          V. Burwitz, 
          \inst{1}
          E. Ferrero, 
          \inst{4}
          \and
          S. Mathur\inst{5}
          }

   \offprints{L. C. Gallo}

   \institute{Max-Planck-Institut f\"ur extraterrestrische Physik, 
Postfach 1312, 85741 Garching, Germany 
         \and
             SRON National Institute for Space Research Sorbonnelaan 2, 3584 CA Utrecht, The Netherlands 
         \and
	     Astronomical Institute, Utrecht University, P.O. Box 80000, 3508 TA, Utrecht, The Netherlands 
	\and
	     Landessternwarte Heidelberg, K\"onigstuhl 12, D-69117, Heidelberg, Germany 
	\and
Department of Astronomy, The Ohio State University, 140 West 18th Avenue,Columbus, OH 43210, USA 
             }

   \date{Received September 15, 1996; accepted March 16, 1997}

   \abstract{
\mrk\ exhibits optical properties of both narrow- and broad-line Seyfert 1
galaxies.  We examine the X-ray properties of this borderline object utilising
proprietary and public data from \chandra, \asca, \rosat\ and \xte, spanning
more than twelve years.  Though long-term flux variability from the pointed
observations appears rather modest ($\sim 3\times$),
we do find examples of rare large amplitude outbursts in the \xte\ monitoring
data.
There is very little evidence of long-term spectral variability as the
low- and high-energy spectra appear constant with time.  A $\sim 6.4\keV$
emission line is detected in the \asca\ spectra of \mrk, but not during the
later, higher flux state \chandra\ observation.
However, the upper limit on the
equivalent width
of a line in the \chandra\ spectrum is consistent with a constant-flux
emission line and a brighter continuum, suggesting that the line is emitted
from distant material such as the putative torus.  
Overall, the X-ray properties of \mrk\ appear typical of BLS1 activity.

   \keywords{galaxies: active --
	     galaxies: individual: \mrk\ --
	     galaxies: nuclei --
	     X-ray: galaxies
               }
   }

   \maketitle
%

\section{Introduction}
\label{sect:intro}

Narrow-line Seyfert 1 (NLS1) galaxies are defined on the bases of
their optical properties, primarily that they possess narrow permitted
lines, weak \oiii\ and strong \feii\
(Osterbrock \& Pogge 1985; Goodrich 1989).
It was realised that these objects often exhibit extreme X-ray
behaviour as well (e.g. Puchnarewicz \et 1992; Boller \et 1996),
such as steep spectra and rapid, large amplitude variability.

However, the traditional definition has become blurred by the discovery
of objects which are not NLS1, but exhibit characteristic X-ray behaviour;
and objects which are NLS1, but do not behave as such (e.g. Reeves \et 2002;
Williams \et 2002, 2004; Dewangan \& Griffiths 2005).
It has been suggested that NLS1 activity is an evolutionary phase
that all active galaxies go through (e.g. Grupe 1996; Mathur 2000),
and there is growing evidence in support of this conjecture.
NLS1 and broad-line Seyfert 1s (BLS1) appear to follow different loci on the
black hole mass -- bulge velocity dispersion plane, implying that black holes
in NLS1 are still growing (Mathur \et 2001; Grupe \& Mathur 2004; Mathur
\& Grupe 2005a, b).
As such, it is important to examine objects which could be in
a transition phase between NLS1 and BLS1.

In consideration of the optical spectrum, \mrk\ is usually defined
as an NLS1.  By virtue of also being a relatively nearby ($z = 0.029$)
PG quasar (PG~0923+129), and X-ray bright in the \rosat\ All-Sky Survey
(RASS; Voges \et 1999), it is often found in many sample studies of AGN
behaviour, but until now it has not benefited from a dedicated examination.

\mrk\ would perhaps be better classified as an intermediate-type NLS1.
The optical properties appear to be on the border that
distinguishes NLS1 from BLS1.  For example,
the $FWHM$ of the H$\beta$ emission line is approximately $1770-2150\kmps$
(e.g. Boroson \& Green 1992; V\'{e}ron-Cetty \et 2001; Botte \et 2004),
and the \oiii\ emission is relatively strong
($\frac{\oiii}{H\beta} \approx 2.4$).
Between $2-10\keV$ the spectrum is somewhat steeper (see Section~\ref{sect:fit})
 than the canonical
$\Gamma \approx 1.9$ found in BLS1 (e.g. Porquet \et 2004)
On the other hand, the low-energy X-ray spectrum,
as observed with \rosat, is relatively flat
and hard\footnote{\label{foot:hr} The hardness ratio ($HR$) used in
this paper will
be defined as $H-S/H+S$, where $H$ and $S$ are count rates in a hard and soft
energy band, respectively.  During the \rosat\ era
$H = 0.4-2.4\keV$ and $S = 0.1-0.4\keV$ were typically used.  We will
adopt this convention unless stated otherwise (as in Sect.~\ref{sect:stime}).}
compared to NLS1 (e.g. Wang \et 1996; Grupe 1996; this study), but typical
for BLS1.
\mrk\ was the only NLS1 that exhibited a
significant hard spectrum at the time Grupe (1996) selected the soft X-ray
AGN sample.

In comparing the known properties of \mrk\ with a sample of NLS1 and BLS1
(Grupe 2004) we find several discrepancies.
\mrk\ has a flat $0.1-2.4\keV$ spectral index more typical
of BLS1.  Like NLS1, it does possess strong \feii\ emission (in terms of
equivalent width), but with respect to H$\beta$ the relative \feii\ emission is
actually weak (comparable
to BLS1).  Finally, as with the H$\beta$ emission line width,
$L/L_{Edd}$ (Baskin \& Laor 2004) lies in the region occupied by
both NLS1 and BLS1.

We report the first dedicated \chandra\ X-ray observation of \mrk, obtained
through the Guaranteed Time programme.  In addition we make
use of archival \rosat, \asca\ and \xte\ data to examine the X-ray
($0.1-10\keV$) properties of \mrk\ in detail.

\section{Observations and data reduction}
\label{sect:data}
X-ray spectroscopic observations of \mrk, which are either proprietary to us or
publicly available, have been performed with \rosat, \chandra\ and
\asca.  The spectroscopic data utilised in this analysis
are described in Table~\ref{tab:log}.

In addition, \mrk\ is observed nearly daily with the \xte\
All-Sky Monitor (ASM).  Quick-look results provided by the ASM/RXTE team
are briefly discussed in Sect.~\ref{sect:time}.

\begin{table}
\caption{X-ray spectroscopy of \mrk.
The telescope and instruments used are given in columns (1) and (2),
respectively.  The start date of the observation is shown in column (3).
Usable energy range, exposure and total counts (source and background)
are given in the following columns.
}
\centering
\begin{tabular}{@{}l|ccccc@{}}
\hline
(1) & (2) & (3) & (4) & (5) & (6) \\
Observatory & Inst. & Date & Energy & Exp. & Counts \\
&  & year.mm.dd & (keV) & (s) & (cts) \\
\hline
\rosat\ & PSPC & 1992.11.05 & 0.1--2.4  & 3855 & 5510 \\
\hline
\rosat\ & PSPC & 1993.26.04 & 0.1--2.4  & 5245 & 5652 \\
\hline
\asca\ & SIS0 & 1998.14.11 & 0.7--10  & 23904 & 13746 \\
& SIS1  &  & 0.7--10  & 20844 & 9536 \\
& GIS2 &  & 1--10  & 29068 & 7871 \\
& GIS3 &  & 1--10  & 29068 & 9242 \\
\hline
\chandra\ & 0-order  & 2004.18.03 & 0.5--7.5 & 20893 & 3545 \\
ASIS-S & MEG &  & 0.6--5 & 20991 & 5989 \\
       & HEG &  & 0.8--7.5 & 20991 & 2559 \\
\hline
\label{tab:log}
$$
\end{tabular}
\end{table}


\subsection{\chandra\ observation}
\mrk\ was observed with \chandra\ on 2004 March 18 with the
Advanced CCD Imaging Spectrometer (ACIS-S) for about $21\ks$.
To minimise pile-up effects on the ACIS image, a grating was inserted in the
photon path; the High-Energy Transmission Grating Spectrometer (HETGS) was
used to disperse photons and so to reduce the count rate on the zeroth order
image
Unfortunately, pile-up was still problematic at about the $30\%$ level.
Accordingly, a pile-up model (Davis 2001) was adopted in all spectral fits.

Data were processed with the Chandra Interactive Analysis of Observations
software (CIAO 3.2), using the Chandra Calibration Database (CALDB 3.0.0).
We applied the recently released, time--dependent gain
correction\footnote{http://asc.harvard.edu/ciao/threads/acistimegain/},
which is necessary to adjust the ``effective gains'', which 
drift with time due to an increasing charge transfer inefficiency (CTI).

The data were filtered to include only the standard event grades
0, 2, 3, 4 and 6. We filtered time intervals
with high background by performing a $3\sigma$ clipping of the
background level using the script
{\tt analyze\_ltcrv}\footnote{http://cxc.harvard.edu
/ciao/threads/filter\_ltcrv/}. This yielded an effective
exposure time of about 21 ks for the ACIS--S3 chip in the
energy range $0.3-10$ keV.

The background was selected from an off--source circular region of radius
$50\arcs$.  To extract the zeroth and higher order grating spectra we
followed the standard procedures.  We analyse the 0-order
spectrum as well as the combined (negative and positive) 1st-order
spectra from the Medium Energy Grating (MEG) and the High Energy Grating (HEG).

\subsection{\asca\ observation}

\asca\ observed \mrk\ for a total of $83\ks$ during which time all
instruments were functioning normally.  The processed data from
all the \asca\ instruments (SIS0, SIS1, GIS2, GIS3) were obtained from the
{\tt TARTARUS} website\footnote{http://astro.ic.ac.uk/Research/Tartarus/}
(version 3.0), where all the
information describing the processing procedures can be found.
The total exposure utilised for this observation is between $20-30\ks$
depending on the instrument (see Table~\ref{tab:log}).

\subsection{\rosat\ observations}

\mrk\ was observed with \rosat\ on numerous occasions as part of the
RASS or because it fell in the field-of-view of other pointings.
Two dedicated observations of \mrk\ with the
Position Sensitive Proportional Counter (PSPC; Pfeffermann \et 1987)
on \rosat\ were performed about one year apart
with the AGN viewed on-axis.  Details of the observations
are found in Table~\ref{tab:log}.

The data were processed in the standard manner with
{\tt EXSAS v03OCT} (Zimmermann \et 1994).   Spectra were created
from source plus
background counts extracted from a circular region with a radius of
2 arcmin and centred on the source.  Background counts were extracted
from a larger off-source region and appropriately scaled to the source
cell size.  Spectral files were then converted into FITS format so that the
data could be analysed in {\tt XSPEC}.

\section{X-ray coordinates}
\label{sect:xcoor}

\begin{figure}
\rotatebox{0}
{\scalebox{0.32}{\includegraphics{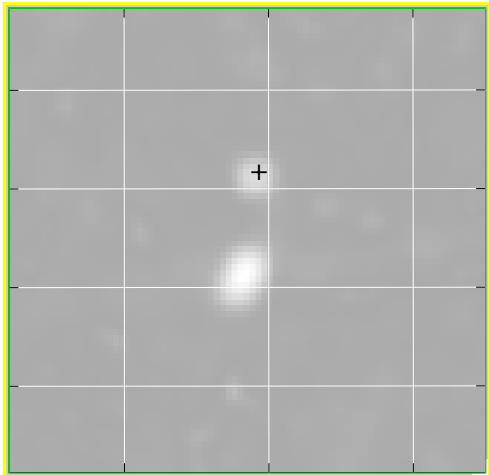}}}
\caption{The cross marks the \chandra\ X-ray position of \mrk\
over-plotted on the \first\ image.  Right ascension ($\alpha$) increases in the
negative
direction along the abscissa.  Declination ($\delta$) increases upward along the
ordinate.  The grid is divided into $30\arcs$ in declination and $3\s$ in
right ascension.
The elongated radio source located about $30\arcs$ from \mrk\ is not
detected with \chandra\ in the X-rays.
}
\label{fig:first}
\end{figure}


The high-resolution capabilities of \chandra\ allows us to determine the
X-ray position of \mrk\ with arcsecond precision.  The X-rays originate
from a point source with coordinates 
$\alpha = 09^{\rm h}26^{\rm m}03^{\rm s}.3$ and
$\delta = +12\deg44^{\prime}04^{\prime\prime}.2$, which agree well with the 
optical position of the
AGN (Clements 1981).  In comparing the \chandra\ image with the PSF
expected from the observation, we found no extended X-ray emission
within the spatial resolution of the instrument
($> 800\pc$ at the redshift of the source).

In Figure~\ref{fig:first} the \chandra\ coordinates
are over-plotted on the archival $20\cm$
\first\footnote{http://sundog.stsci.edu/} (White \et 1997)
image obtained with the VLA.  Although \mrk\ is considered radio-quiet
it is detected in the sensitive \first\ survey with an integrated flux 
density of about $8.5\mJy$.
The X-ray position is also in good agreement with the origin of the radio
emission.

Of interest may be the elongated radio source located $\sim 30\arcs$ south
of \mrk.  In the low-resolution instruments (i.e. \rosat\ and \asca),
where the source extraction region is typically arcminutes in size, this source
could contribute to the X-ray flux of \mrk.  Little is known about this
object other than that it is associated with the infrared source
2MASX~J09260351+1243341, with a K-band flux density of $2.52\pm0.40\mJy$,
and a \first\ ($1.4$GHz) integrated flux density of
$27.5\pm0.37\mJy$.

At the redshift of \mrk, 2MASX~J09260351+1243341 would be located
$\approx24\kpc$ from the AGN.  We considered the possibility that this could
be a radio jet associated with \mrk.
However, the infrared emission, assuming a jet with a power-law
spectrum ($F_\nu \propto \nu^{-\alpha}$) and typical spectral index in the
range $0.5-0.8$
(e.g. Ferrari 1998; Worrall \& Birkinshaw 2004), is at least $25$ times too
high.
More importantly, the object is not detected in the
$\sim 20\ks$ \chandra\
exposure, indicating that the X-ray fluxes measured with \asca\ and \rosat\
are dominated by \mrk.

\section{Spectral analysis}
\label{sect:fit}

The energy band in which the data from each instrument is analysed is
given in Table~\ref{tab:log}.
All of the spectra were grouped such that each bin contained at least 20
counts. Spectral fitting was performed with {\tt XSPEC v11.3.1} (Arnaud
1996) and fit parameters are reported in the rest frame of the object.
The quoted errors on the model parameters correspond to a 90\% confidence
level for one interesting parameter (i.e. a $\Delta\chi^2 = 2.7$ criterion).
K-corrected luminosities were derived assuming isotropic emission, a
value for the Hubble constant of $H_0$=$\rm 70\ km\ s^{-1}\ Mpc^{-1}$, and
a standard cosmology with $\Omega_{M}$ = 0.3 and $\Omega_\Lambda$ = 0.7.
The value of the Galactic column density toward \mrk\ was taken to be
$\nh = 4.03\times 10^{20} \pscm$ (Elvis \et 1989).

\subsection{The \rosat\ spectra}
\label{sect:rosat}

The PSPC is sensitive in the $0.1-2.4\keV$ range and this offers the opportunity
to examine lower levels of absorption than possible with \asca\ or
\chandra.
First attempts were made to fit the \rosat\ spectra of \mrk\ with a power-law
modified
by a neutral hydrogen column.  Leaving the absorption component free to
vary resulted in measurements consistent with the value
given by Elvis \et (1989).  Fixing the photoelectric absorption to the Galactic
value and including an additional component to estimate cold absorption
intrinsic to the source did not enhance the fits or result in any
meaningful detection of intrinsic absorption.
As such, in subsequent fits (in particular with the
\asca\ and \chandra\ data) the photoelectric absorption component was simply
fixed to the Galactic value.

\begin{figure}
\rotatebox{270}
{\scalebox{0.32}{\includegraphics{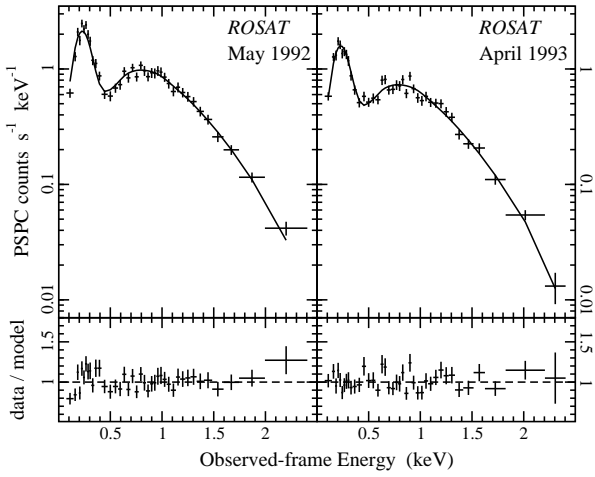}}}
\caption{The PSPC count rate spectrum fitted with a power-law with
Galactic absorption (upper panels).  The fit ratios are shown in the
lower panels.  The spectrum from the 1992 observation is shown on the left
and from the 1993
observation on the right.
}
\label{fig:rospo}
\end{figure}

A simple power-law was a reasonable fit to the 1993 (second) observation
of \mrk\ (see Figure~\ref{fig:rospo} and Table~\ref{tab:rosfit}).
Including additional components
to mimic a multi-component continuum did not improve the simple
model.  However, the 1992 (first) PSPC observation appeared more
complicated.  A simple power-law provided a mediocre fit ($\chi^{2}_{\nu}/$dof $=
1.50/38$) to the $0.1-2.4\keV$ data.  Multi-component models
(e.g. broken power-law, and blackbody plus power-law) provided a substantial
improvement over the single power-law (Table~\ref{tab:rosfit}).

The most apparent difference between the two \rosat\ observations is that
the 1992 spectrum requires a second component in addition to a power-law.
However, the hardness ratio at both epochs
is comparable ($HR_{1992} = 0.29\pm0.07$, $HR_{1993} = 0.29\pm0.03$),
indicating that the spectral differences between the two observations
may not be as significant as suggested by the modelling.  Indeed, the
primary differences in the two spectra are between $2-2.4\keV$,
where the sensitivity
of the instrument begins to diminish.  Therefore, it is not conclusive whether
apparent changes are due to intrinsic variability in the AGN or
signal-to-noise.

Results from modelling the \rosat\ PSPC data of \mrk\ are tabulated
in Table~\ref{tab:rosfit}.
\begin{table}
\caption{\rosat\ PSPC ($0.1-2.4\keV$) spectral fit results.
The model and fit parameters are given in columns (1) and (2),
respectively.  The values of the parameters during the first (1992) and
second (1993) observation are given in columns (3) and (4), respectively.
Unabsorbed flux ($F$) and luminosity ($L$) are given for the power-law
fit in units of $10^{-11}\ergpscmps$ and $10^{43}\ergps$, respectively.
The values marked with $u$ indicate that the parameter is unconstrained and
accurate uncertainties could not be estimated.
}
\centering
\begin{tabular}{@{}l|ccc@{}}
\hline
(1) & (2) & (3) & (4)  \\
Model & Parameters & Obs 1 & Obs 2   \\
&  & (1992) & (1993)  \\
\hline
power-law & $\chi^{2}_{\nu}$/dof & $1.50/38$ & $1.20/38$ \\
          & $\Gamma$ & $2.49\pm0.04$ & $2.51\pm0.04$ \\
          & $F$               & $3.81$  & $2.85$   \\
          & $L$               & $7.45$  & $5.58$   \\
\hline
broken    & $\chi^{2}_{\nu}$/dof & $1.41/36$ & $1.24/36$ \\
power-law & $\Gamma_1$ & $2.61\pm0.06$ & $2.52\pm0.03$ \\
          & $E_b$ (keV)& $0.61\pm0.13$ & $1.70^{u}$ \\
          & $\Gamma_2$ & $2.33\pm0.08$ & $1.50^{u}$ \\
\hline
blackbody & $\chi^{2}_{\nu}$/dof & $1.34/36$ & $1.25/36$ \\
plus      & $kT$ (eV) & $62^{+13}_{-20}$ & $80^{+25}_{-36}$ \\
power-law & $\Gamma$  & $2.29^{+0.06}_{-0.20}$ & $2.45\pm0.04$ \\
\hline
\label{tab:rosfit}
\end{tabular}
\end{table}

\subsection{The \asca\ spectra}
\label{sect:asca}

Modelling the data from the four \asca\ instruments separately revealed
relatively good agreement in the measured fit parameters (within 90 per cent
confidence levels).  As such, the data from all the instruments were fitted
simultaneously while allowing for an energy-independent scale factor between
the SIS and GIS spectra.  The residuals
from each instrument were examined separately to ensure agreement.
Since there was no evidence of intrinsic cold absorption from the \rosat\
observations, the photoelectric absorption parameter was fixed to the value
of the Galactic column.

\begin{figure}
\rotatebox{270}
{\scalebox{0.32}{\includegraphics{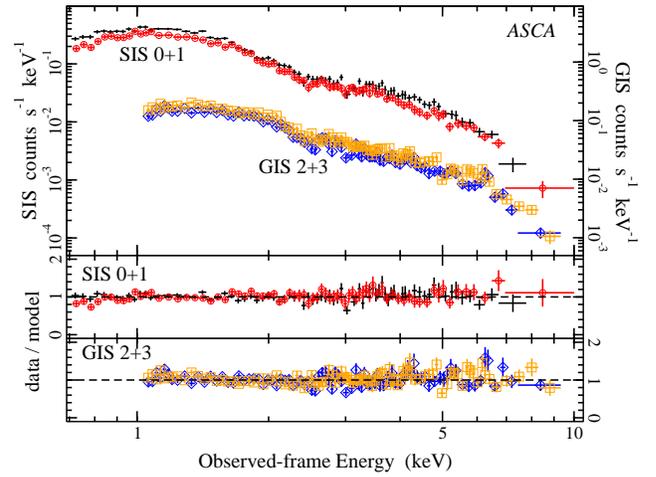}}}
\caption{Top panel:  \asca\ spectral data from the SIS0 (black crosses)
and SIS1 (red open circles) instruments, and the two GIS instruments
(blue diamonds and orange squares correspond to
GIS2 and GIS3, respectively).  The GIS flux density is given on the
right axis.  Middle panel:  SIS residuals from the best-fit power-law plus
Gaussian model (see text and Table~\ref{tab:ascafit} for details)
fitted simultaneously with all \asca\ instruments.
Lower panel: Same as middle panel, but for the GIS instruments.
The data are binned for display purposes.
}
\label{fig:ascafit}
\end{figure}

Fitting the $2.5-10\keV$ spectra with a power-law ($\Gamma \approx 1.96$)
resulted in a reasonable
fit ($\chi^2_{\nu}/$dof $= 1.12/239$); however positive residuals remained at
$\sim 6.3\keV$ in the observed-frame.  Adding a Gaussian profile to the
power-law model improved the fit considerably ($\Delta\chi^2 = 21.6$ for 3
additional free parameters), suggesting emission from Fe~\ka\ at a rest frame
energy of $E \approx 6.45\keV$.  Extrapolating
the fit to 0.7\keV\ in the SIS and 1\keV\ in the GIS
showed some residuals at lower energies.  These residuals did not require
additional model components to minimise, but could be adequately
improved by simply allowing for a slight change in the slope of the
current power-law.
The resulting unabsorbed $2-10\keV$ flux was
$F \approx 1.1 \times 10^{-11}\ergpscmps$, and the corresponding intrinsic
luminosity was $L \approx 2.1 \times 10^{43}\ergps$.
The best-fit model is displayed in Figure~\ref{fig:ascafit}
and the parameters are given in Table~\ref{tab:ascafit}.
\begin{table}
\caption{Best-fit power-law plus Gaussian profile for the \asca\ data.
The \asca\ instruments used in each fit are given in column (1), and
the quality of the fit is shown in column (2).
The power-law photon index is provided in column (3).
The line parameters energy, line width, and equivalent width are given in
columns (4), (5), and (6), respectively.
}
\centering
\label{tab:ascafit}
\begin{tabular}{@{}l|ccccc@{}}
\hline
(1) & (2) & (3) & (4) & (5) & (6) \\
Instrument & $\chi^2_\nu$/dof & $\Gamma$ & $E$ & $\sigma$ & $EW$ \\
 &  &  & (keV) & (eV) & (eV) \\
\hline
SIS0+1 & 0.90/175 & $2.02\pm0.03$ & $6.63^{+0.17}_{-0.35}$ & $<415$ & $305^{+18}_{-32}$ \\
\hline
GIS2+3 & 1.18/209 & $2.10\pm0.03$ & $6.33^{+0.12}_{-0.17}$ & $<491$ & $420^{+40}_{-42}$\\
\hline
SIS+GIS & 1.09/384 & $2.06\pm0.02$ & $6.36^{+0.14}_{-0.17}$ & $<576$ & $384^{+28}_{-20}$\\
\hline
\end{tabular}
\end{table}

\subsection{The \chandra\ spectra}
\label{sect:chan}
In order to minimise the effects of pile-up in the zeroth order
spectrum, the pile-up model of Davis (2001) was used in all spectral
fits.  The parameters of this model component are give in
Table~\ref{tab:chanfit}.

\begin{figure}
\rotatebox{270}
{\scalebox{0.32}{\includegraphics{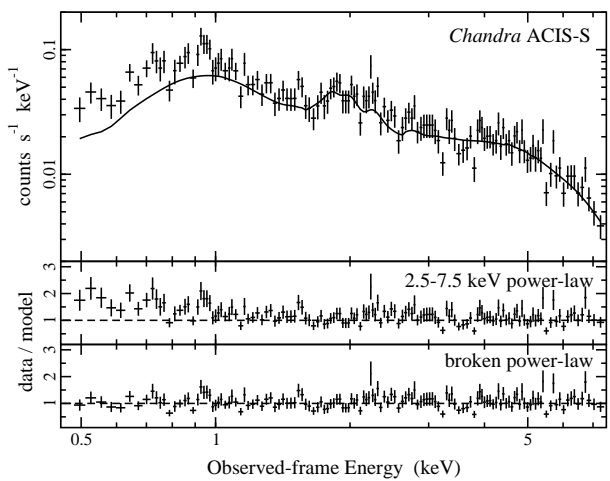}}}
\caption{The \chandra\ $0.5-7.5\keV$ count spectra, fitted with a
$2.5-7.5\keV$ power-law extrapolated to lower-energies, is shown in the
top panel.  The ratio from this fit is shown in the middle panel.  Note
the excess below $\sim 1\keV$.  In the bottom panel are the residuals
(data/model) from fitting the spectra with a broken power-law
(see text and Table~\ref{tab:chanfit} for details).
The data are binned for display purposes.
}
\label{fig:chanfit}
\end{figure}

As with the \asca\ analysis, we began by fitting the high-energy
($2.5-7.2\keV$) \chandra\ (0-order)
data with a power-law ($\Gamma = 2.10^{+0.36}_{-0.34}$) modified by Galactic
absorption.
This resulted in a rather poor fit ($\chi^2_\nu/$dof $= 1.24/61$), and there were
no indication of possible emission features (Figure~\ref{fig:chanfit}).

Extrapolating the power-law down to $0.5\keV$, showed evidence of a slight
excess below $\sim 1\keV$ (Figure~\ref{fig:chanfit}).
Fitting the broadband continuum with just the single power-law (as was done
to the \asca\ data) did not result in an entirely satisfactory fit
(see Table~\ref{tab:chanfit} for broadband fit results).

The addition of a second continuum component, in the form of a blackbody
or a broken power-law, improved the power-law model equally
well ($\Delta\chi^2 = 8.3$ for 2 additional free parameters).
The unabsorbed $0.5-2\keV$ and $2-10\keV$ fluxes obtained from the broken
power-law fit were $2.22 \times 10^{-11}$ and
$2.31 \times 10^{-11}\ergpscmps$, respectively.
The $2-10\keV$ rest frame luminosity was $4.47 \times 10^{43}\ergps$.

The addition of a Gaussian profile did not significantly improve the
fit ($\Delta\chi^2 = 1.5$ for 3 additional parameters).  Fixing a narrow
($\sigma = 1\eV$), Gaussian profile to $6.4\keV$, as was found with
\asca, allowed estimation of a $90\%$ upper-limit on the equivalent
width of $EW < 149\eV$.

The best-fit models to the $0.5-7.5\keV$ \chandra\ data are described
in Table~\ref{tab:chanfit}.
The available statistics in the HEG and MEG are quite low.  Nevertheless,
application of the zeroth order best-fit model to the grating spectra
resulted in comparable fits.

\begin{table}
\caption{Best-fit models to the \chandra\ data.  In column (1) the
adopted continuum model is given, and in column (2) the quality of fit.
In column (3) the continuum model parameters, and in column (4)
the parameters from the pile-up model are shown. The grade morphing
($\alpha$) and the PSF fraction treated for pile-up ($psf$) are free
parameters (see Davis 2001 for details).
}
\centering
\label{tab:chanfit}
\begin{tabular}{@{}l|ccc@{}}
\hline
(1) & (2) & (3) & (4) \\
Continuum & $\chi^2_\nu$/dof & Continuum & Pile up \\
Model &  & Parameters & Parameters \\
\hline
power-law & 1.26/137 & $\Gamma = 2.24\pm0.21$ & $psf = 0.91^{+0.04}_{-0.09}$ \\
          &          &                        & $\sigma = 0.43^{+0.45}_{-0.10}$ \\
\hline
broken    & 1.22/135 & $\Gamma_1 = 2.52^{+0.28}_{-0.30}$ &  $psf = 0.89^{+0.05}_{-0.30}$ \\
power-law &          & $E_{b} = 1.17^{+0.74}_{-0.40}\keV$ & $\sigma = 0.32^{+0.68}_{-0.12}$ \\
          &          & $\Gamma_2 = 1.96\pm0.20$ & \\
\hline
blackbody & 1.22/135 & $kT = 105\pm23\eV$ & $psf = 0.89^{+0.05}_{-0.23}$ \\
plus      &          & $\Gamma = 1.96\pm0.23$ & $\sigma = 0.34^{+0.66}_{-0.12}$ \\
power-law &          &                        &  \\
\hline
\end{tabular}
\end{table}


\section{Timing properties}
\label{sect:time}

\subsection{Long-term timing behaviour}
\label{sect:ltime}
\begin{figure}
\rotatebox{270}
{\scalebox{0.32}{\includegraphics{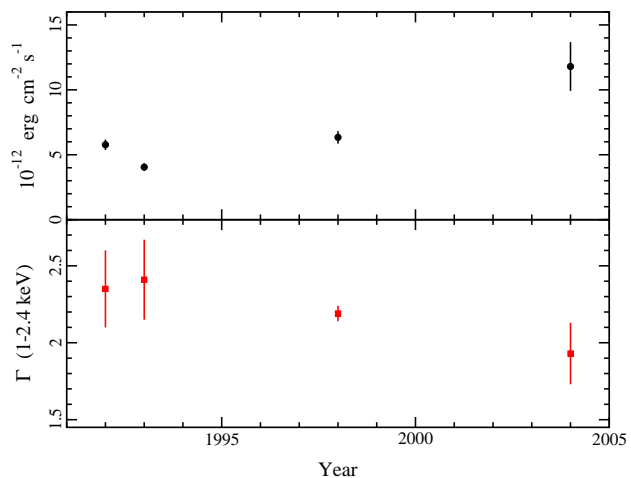}}}
\caption{Flux (upper panel) and spectral (lower panel) variability curves
built from the four pointed observations utilised in this analysis.
The energy range used to make the measurements is $1-2.4\keV$ because it is the
only band sampled by all the instruments.
The photon index ($\Gamma$) is obtained from fitting
the $1-2.4\keV$ range with a power-law, and the error bars in the flux
are estimated from the uncertainties in the normalisation
of the power-law at $1\keV$.
}
\label{fig:multilc}
\end{figure}

The four observations analysed here were spread over more than ten years;
however the flux in the $1-2.4\keV$ range, which is the only band sampled
at every epoch, varied by less than a factor of three
(see Figure~\ref{fig:multilc}).
The variability appears modest in comparison
to other Seyferts (e.g. Markowitz \& Edelson 2004), but a fair comparison
cannot be made given our poorly sampled long-term light curve.

Moreover, there is no significant spectral variability over this period.
Within $90\%$ uncertainties, the spectral slope between $1-2.4\keV$
is constant (Figure~\ref{fig:multilc}).
In addition, the spectral fits to the two \rosat\ and
\chandra\ data sets suggest that the low-energy
($E < 1\keV$) spectral form remains constant, and comparable to a power-law
with a spectral index of $\Gamma \approx 2.5$.

\begin{figure}
\rotatebox{270}
{\scalebox{0.32}{\includegraphics{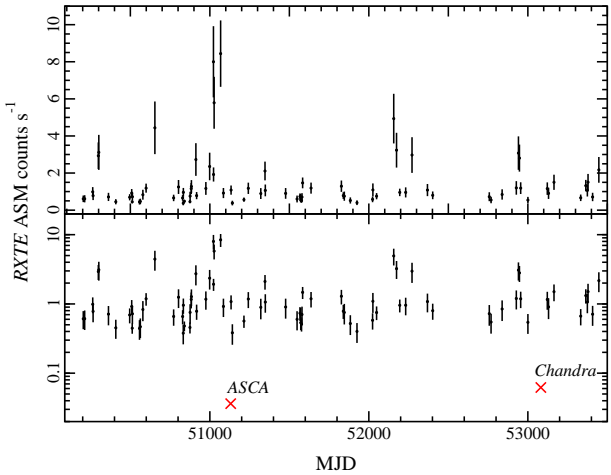}}}
\caption{Long-term \xte\ ASM light curve constituting only data when
\mrk\ was detected above the $3\sigma$ level.  The light curve is shown
in linear (top panel) and logarithmic (bottom panel) \cps.
The red X mark the estimated ASM count rate of \mrk\ during the \asca\ and
\chandra\ observations.
}
\label{fig:rxtelc}
\end{figure}

\mrk\ has been observed nearly daily with the \xte\ ASM since 1996 January.
As of 2005 April 27 over 2600 observations were made.  \mrk\ has been detected
above a $3 \sigma$ level about 85 times.  In Figure~\ref{fig:rxtelc}, the
$> 3 \sigma$ data points from the quick-look $2-10\keV$ light curve provided
by the ASM/RXTE team are plotted.
Using the \asca\ and \chandra\ best-fit models, we have estimated an ASM
count rate with
{\tt WebPIMMS}\footnote{http://heasarc.gsfc.nasa.gov/Tools/w3pimms.html}, and
over-plotted these estimates
on Figure~\ref{fig:rxtelc}.
Since most of the time \mrk\ is not significantly detected, it seems
reasonable to assume that during the \asca\ and \chandra\ observations
(and likely the \rosat\ observations as well), \mrk\ was in a typical
flux state.  Assuming so, the ASM light curve demonstrates several examples
of flaring events in which the $2-10\keV$ flux increases by one, possibly
even two orders of magnitude.

\subsection{Short-term timing behaviour}
\label{sect:stime}

\begin{figure}
\rotatebox{270}
{\scalebox{0.32}{\includegraphics{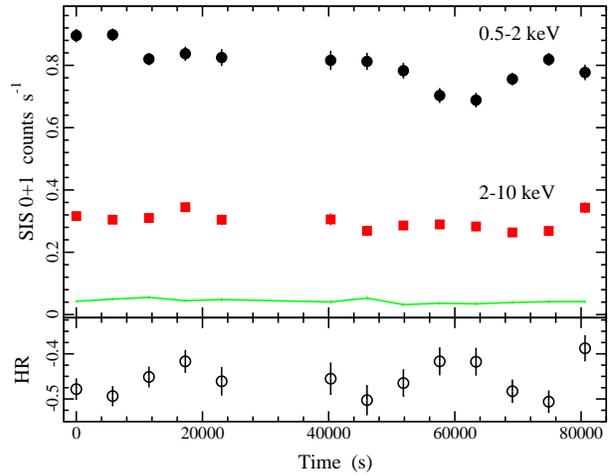}}}
\caption{Top panel: \asca\ soft ($0.5-2\keV$) and hard ($2-10\keV$)
combined SIS light curves in $5760\s$ bins.
The background light curve in the $0.5-10\keV$ band is marked by the
green solid line.
Lower panel: The hardness ratio ($HR$, see footnote~\ref{foot:hr}),
is plotted
against elapsed time.  In this
case, $H$ and $S$ are as defined in the top panel.
}
\label{fig:ascalc}
\end{figure}

The combined \asca\ SIS0 and SIS1 light curves in a hard ($2-10\keV$)
and soft ($0.5-2\keV$) band over the duration ($\sim 83\ks$) of the observation
are shown in Figure~\ref{fig:ascalc}.  On this time scale, both light curves
fluctuate by $\pm 10\%$ about the mean and are inconsistent with a constant
according to a $\chi^2$-test ($\chi^2 = 99.0$ and $40.0/12$ dof for the
soft and hard light curves, respectively).  In addition, the hardness
ratio between these two bands also shows variations over this period
(lower panel of Figure~\ref{fig:ascalc}); however, there is no clear
correlation between flux and spectral variability.

When examining the \chandra\ light curve, and the \asca\ light curves during
each telescope revolution, the fluctuations are greatly diminished
(primarily due to statistics).
For example, the $\sim 20\ks$ \chandra\ light curve in $500\s$ bins was
entirely consistent with a constant ($\chi^2_{\nu} < 1$).
To compare the rapid variability in \mrk\ to other AGN we calculated the
excess variance ($\sigma^{2}_{\rm rms}$) following Nandra \et (1997), using
$128\s$ binning of the $0.5-10\keV$ SIS light curve.  We estimate a value
of $\sigma^{2}_{\rm rms} \approx 0.26 \times 10^{-2}$.  In comparing 
the excess variance derived for \mrk\ with figure 3 of Leighly (1999),
in which both NLS1 and BLS1 are plotted, it is clear that the weak variations
of \mrk\ on short time scales are typical of BLS1 of comparable luminosities.

\section{Conclusions}
\label{sect:conc}

We present spectra from four pointed observations of the borderline
NLS1, \mrk, utilising
data from \rosat, \asca\ and \chandra\ which span over ten years.
In addition, we also present the $2-10\keV$ light curve obtained with the
\xte\ ASM.
The main results of this analysis are the following:
\begin{itemize}

\item
The X-ray position is well correlated with the optical and radio
positions of \mrk.  There is an unidentified radio source located
within $30\arcs$ of \mrk; however, this object is not detected
with \chandra, suggesting that the X-ray emission measured with \rosat\
and \asca\ is dominated by \mrk.

\item
Comparing the measured flux during the pointed observations with
the long-term ASM light curve, the four spectra of \mrk\
appear to have been obtained during flux states typical for the source.
The flux during these pointed observations is comparable within a factor
of $\sim3$.

\item
In the \rosat\ and \chandra\ observations there is indication of an
additional low-energy emission component,
which appears constant in spectral form ($\Gamma \approx 2.5$)
at all three epochs.

\item
The high-energy ($2-10\keV$) photon index does not vary significantly between
the \asca\ and
\chandra\ observations, although the flux increases by about a factor of two.
An Fe~\ka\ emission with $EW \approx 380\eV$ is detected in the \asca\
spectrum.  A line is not statistically required in the \chandra\ data, and
the upper-limit on the equivalent width of a narrow $6.4\keV$ line is
$EW<149\eV$.  The measured equivalent widths at the two epochs are in
agreement with the flux of the line remaining constant while the continuum
flux varies.  This suggests that the line probably originates in distant
material such as the torus.

\item
The long-term light curve obtained with the \xte\ ASM indicates that large
amplitude flaring-type events are rare in \mrk, but they do occur.

\item
During the $\sim 83\ks$ \asca\ observation, \mrk\ does show fluctuations
of about $\pm 10\%$ in the hard, soft and hardness ratio variability curves.
On shorter time scales the variability is diminished.

\end{itemize}

The optical properties of \mrk\ place it on the borderline distinguishing
NLS1 and BLS1.  The X-ray properties examined at various epochs over more
than ten years do not allow a clearer definition, but suggest that \mrk\
is more similar to BLS1.
The rather flat slope of the soft-excess, weak short-term variability
and neutral iron line (Porquet \et 2004) are consistent
with BLS1 X-ray properties.  However, there is some evidence from the
\xte\ light curve of, albeit rare, large amplitude fluctuations, which
have been seen frequently in some NLS1 (e.g. Boller \et 1997; Brandt \et 1999).


Deeper studies of samples and individual objects which seem to exhibit
properties of both NLS1 and BLS1 could be fruitful in establishing a more
accurate defining scheme, as well as revealing the underlying physics.


\begin{acknowledgements}
This research has made use of the Tartarus (Version 3.0) database,
created by Paul O'Neill and Kirpal Nandra at Imperial College London, and
Jane Turner at NASA/GSFC. Tartarus is supported by funding from PPARC, and
NASA grants NAG5-7385 and NAG5-7067.
This work has also made use of results provided by the
ASM/RXTE teams at MIT and at the RXTE SOF and GOF at NASA's GSFC.
Many thanks are due to the referee for a quick report.
\end{acknowledgements}

\end{document}